\documentclass[review]{elsarticle}

\usepackage{color,xcolor}
\usepackage{amssymb}

\usepackage{graphicx} 
\usepackage{epstopdf}
\usepackage{ulem}

\begin{document}

\begin{frontmatter}

\title{Solving the Grad-Shafranov equation using spectral elements for tokamak equilibrium with toroidal rotation}

\author{Haolong Li}
\address{CAS Key Laboratory of Geospace Environment and Department of Engineering and Applied Physics,
University of Science and Technology of China, Hefei, Anhui 230026, China}

\author{Ping Zhu\corref{mycorrespondingauthor}}
\cortext[mycorrespondingauthor]{Corresponding author}
\ead{zhup@hust.edu.cn}

\address{International Joint Research Laboratory of Magnetic Confinement Fusion and Plasma Physics, State Key Laboratory of Advanced Electromagnetic Engineering and Technology, School of Electrical and Electronic Engineering, Huazhong University of Science and Technology, Wuhan, Hubei 430074, China}

\address{Department of Engineering Physics, University of Wisconsin-Madison, Madison, Wisconsin 53706, USA}

\begin{abstract}
The Grad-Shafranov equation is solved using spectral elements for tokamak equilibrium with toroidal rotation. The Grad-Shafranov solver builds upon and extends the NIMEQ code [Howell and Sovinec, Comput. Phys. Commun. 185 (2014) 1415] previously developed for static tokamak equilibria. Both geometric and algebraic convergence are achieved as the polynomial degree of the spectral-element basis increases. A new analytical solution to the Grad-Shafranov equation is obtained for Solov'ev equilibrium in presence of rigid toroidal rotation, in addition to a previously obtained analytical solution for a defferent set of equilibrium and rotation profiles. The numerical solutions from the extended NIMEQ are benchmarked with the analytical solutions, with good agreements. Besides, the extended NIMEQ code is benchmarked with the FLOW code [L. Guazzotto, R. Betti, et al., Phys. Plasma 11(2004)604].
\end{abstract}


\end{frontmatter}


\section{Introduction}
\label{sec:intro}

For the static equilibrium, the magnetohydrodynamic (MHD) equations yield nonlinear second order differential equation known as Grad-Shafranov equation\cite{grad1958,shafranov1958}. The steady state equilibria defined by the solutions of Grad-Shafranov (GS)\cite{grad1958,shafranov1958} equation act as the foundation for evaluating the MHD stability of tokamak plasma. Numerical codes have been developed based on different algorithms to solve nonlinear GS equation for given plasma density, temperature and magnetic field profiles directly from experiment\cite{chease,efit}. However, most of these codes have only considered the static tokamak equilibrium where plasma flow such as the toroidal rotation is absent.

Toroidal rotation plays significant roles in many tokamak plasma processes. For example, plasma flow and flow shear above certain threshold may lead to the formations of H-mode and internal transport barrier (ITB)\cite{LH_ITB_influence1,LH_ITB_influence2,LH_ITB_influence3}. Meanwhile, plasma flow and flow shear can also directly affect plasma stability and transport\cite{FLOW_code,flow_observ_EAST,NTM_influence3,RWM_influence6,ELM_influence2,yanxt2017}. In particular, flow shear may have stabilizing effects on neoclassical tearing modes(NTMs)\cite{NTM_influence1,NTM_influence2}, tearing modes (TMs)\cite{TM_influence1,TM_influence2,TM_influence3,TM_influence4} and edge localized modes(ELMs)\cite{ELM_influence1,ELM_influence2,ELM_influence3,ELM_influence4}.  It is found that sufficient toroidal flow opens up a  stability window for resistive wall mode (RWM)\cite{RWM_influence1,RWM_influence2,RWM_influence4,RWM_influence5,RWM_influence6}. On the other hand, plasma flow and shear can also directly modify plasma equilibrium due to the centrifugal effect.

There is a rich history of analytic solution to the GS equation\cite{solov_analy_solution1,solov_analy,mc1999analytical,analy_review2014,FRC_eq2003,FRC_eq1981,wangSJ2004}. For example, the solution of the GS homogeneous equation is given by S. B. Zheng\cite{solov_analy_solution1}. The inhomogeneous GS equation with linear source function $P$ and $F$ known as  Solov$'$ev equilibrium can be solved analytically for any two parameters\cite{solov_analy_solution1,solov_analy}. The solution to the GS equation with parabolic source functions has been also reported, which allow independent specifications of plasma current density, pressure ratio and one shape moment such as the internal inductance\cite{mc1999analytical,analy_review2014}.
Besides tokamak, equilibria of other configurations also haved been obtained analytically, such as those of the field-reversed configuration (FRC)\cite{FRC_eq2003,FRC_eq1981}.

However, the equilibriums that can be described using analytic solutions of GS equation are limited. GS equation often has to be solved numerically, based on the choice of either the flux along boundary or the source functions. Fixed-boundary solvers specify the flux value along the boundary of computation domain. Free-boundary solvers self-consistantly calculate the flux value along the boundary of computation, combining the contribution from external magnetic coils and the contribution from internal plasma current. Various numerical methods have been applied to solving the GS equation, for example, finite difference\cite{finit_difference1979}, spectral methods\cite{spectral1985}, Green’s functions\cite{green1985}, linear finite elements\cite{finit_element1981,finit_element1987}, and Hermite cubic finite elements\cite{bicubic_Hermite_element1992}. Consequently, many numerical toroidal equilibrium codes have been developed, such as EFIT\cite{efit}, CHEASE\cite{chease}, ESC\cite{esc1999}, NIMEQ\cite{nimeq}, etc.

 In addition, several codes are able to solve for toroidal equilibrium in presence of flow, such as FLOW\citep{FLOW_code}, CLIO\cite{clio_code} and FINESSE\cite{finesse_code}. But these codes are often designed for topologically toroidal domains and do not consider the regularity issues associated with
the $R^{-1}$ singularity, where $R$ is the major radius. This issuse would arise in topologically cylindrical domains, which include the geometric axis $R=0$.

Previously, a Grad-Shafranov solver NIMEQ\citep{nimeq} was developed for static toroidal equilibrium within the framework of NIMROD\cite{nimrod}. In this work, we extend the Grad-Shafranov solver NIMEQ\citep{nimeq} to solution of the toroidal equilibrium in presence of toroidal rotation. A new analytical solution of the modified Grad-Shafranov equation is found. The extended NIMEQ is benchmarked with the new analytical solution and the analytical solution by Maschke and Perrine\citep{M_P}. The convergence of the extended NIMEQ is tested with h-refinement and p-refinement methods. Furthermore, the extended NIMEQ is benchmarked with FLOW in a convergence study.

The rest of this paper is organized as follows. Section (\ref{sec:model}) reviews the Grad-Shafranov equation with toroidal rotation. Section (\ref{sec:analy_solution}) shows  a new analytical solution to the modified Grad-Shafranov equation along with the analytical solution obtained by Maschke and Perrin\citep{M_P}. Section (\ref{sec:numer}) presents the numerical algorithm of the extended NIMEQ. Benchmarking and convergence studies are performed with these two equilibria in section (\ref{sec:bench_converg}). Finally, section (\ref{sec:conclusion}) gives conclusion and discussion.

\section{Grad-Shafranov equation with toroidal rotation}
\label{sec:model}

Tokamak equilibria with toroidal rotation are governed by four equations: the force balance equation, magnetic divergence constraint, Ampere's law and state equation of ideal gas\citep{fm_flow}

\begin{eqnarray}
\rho(\vec{u}\cdot \nabla)\vec{u}=-\nabla P+\vec{J} \times \vec{B}
\label{intro:force_balance}
\\
\nabla \cdot \vec{B} =0
\label{intro:mag_divg}
\\
\mu_{0} \vec{J} = \nabla \times \vec{B}
\label{intro:ampere}
\\
P= \frac{\rho}{m_{i}} T
\label{intro:ideal_gas}
\end{eqnarray}
where $\vec{u}=R^{2} \Omega \nabla \phi$ denotes the toroidal flow velocity, $\Omega$ the frequency of toroidal rotation, $P$ the plasma pressure, $\vec{J}$ the plasma current density, $\vec{B}$ the magnetic field and $\mu_{0}$ the permeability of vacuum. Besides, $\rho$ denotes the mass density, defined as $\rho \equiv m_{i} n_{i}+ m_{e} n_{e} \simeq m_{i}n$, $n \equiv n_{i}=n_{e}$ and $T$ denotes the plasma temperature defined as $T \equiv T_{i}+T_{e}$, where $m_{i}$($m_{e}$), $n_{i}$($n_{e}$) and $T_{i}(T_{e})$ are the ion (electron) mass, number density and temperature.

The magnetic field is expressed as $\vec{B}=\nabla \phi \times \nabla \psi +F\nabla \phi$ and the plasma current is expressed as $\mu_{0}\vec{J}=\mu_{0}RJ_{\phi} \nabla \phi + \nabla F \times \nabla \phi$ in the cylindrical coordinate system and $F(\psi)=RB_{\phi}$ is a flux function \cite{nimeq}. From the curl of Ohm's law, it is observed that the frequency of toroidal rotation is a flux function $\Omega=\Omega(\psi)$. Substituting these above expressions for $\vec{B}$, $\vec{J}$ and $\vec{u}$ into Eq.(\ref{intro:force_balance}) yields:

\begin{eqnarray}
\rho R \Omega^{2} - \frac{\partial P}{\partial R} = 0
\label{intro:R-direc}
\\
\Delta^{*} \psi = - R^{2} \frac{\partial P}{\partial \psi} - F \frac{dF}{d\psi}
\label{intro:psi-direc}
\end{eqnarray}
where the Grad-Shafranov operator is defined as 

\begin{eqnarray}
\Delta^{*} \equiv R \frac{\partial}{\partial R} R^{-1} \frac{\partial}{\partial R} + \frac{\partial^{2}}{\partial Z ^{2}}
\label{intro:delta_star}
\end{eqnarray}

For fusion plasma the thermal conduction along magnetic field lines is fast compared to the heat transport perpendicular to a magnetic surface. Thus, plasma temperature can be considered as a flux function, namely $T=T(\psi)$. From Eq.(\ref{intro:R-direc}), the pressure is integrated as:

\begin{eqnarray}
P(\psi,R)=P_{0}(\psi) \exp \left[ \frac{m_{i} \Omega^{2} R_{0}^{2}}{2T} \left( \frac{R^{2}}{R_{0}^{2}} -1 \right) \right]
\label{intro:pres_flow}
\end{eqnarray}

Substituting $P(\psi,R)$ into Eq.(\ref{intro:psi-direc}), we have

\begin{eqnarray}
\Delta^{*} \psi = - F \frac{d F}{d \psi} -                                                                                                                                                                                                                                                                                                                                                                                                                                                                                                                                                                                                                                                                                                                                                                                                                                                                                                                                                                                                                                                                                                                                                              \mu_{0} R^{2} \left[ \frac{dP_{0}}{d\psi} - P_{0} \frac{m_{i} R_{0}^2 \Omega}{T} (\frac{R^{2}}{R_{0}^{2}}-1) \frac{d\Omega}{d\psi} \right.
\nonumber
\\
\left. + P_{0} \frac{m_{i} R_{0}^{2} \Omega^{2}}{2T^{2}} (\frac{R^{2}}{R_{0}^{2}}-1) \frac{dT}{d\psi} \right] \exp \left[ \frac{m_{i} \Omega^{2} R_{0}^{2}}{2T} \left( \frac{R^{2}}{R_{0}^{2}} -1 \right) \right]
\label{gs_modf_full}
\end{eqnarray}
where $R_{0}$ denotes the position of magnetic axis. $P=P_{0}(\psi)$ when $\Omega=0$. In the limit $\Omega \rightarrow 0$, the static equilibrium pressure can be recovered as a flux function. Meanwhile, Eq.(\ref{gs_modf_full}) will reduce to the static GS equation.

\section{Analytical solutions}
\label{sec:analy_solution}

\subsection{Solov'ev equilibrium with toroidal rotation}
\label{eqs_new}

We obtain a new analytical solution to Eq.(\ref{gs_modf_full}) for Solov'ev equilibrium in presence of toroidal rotation. In Solov'ev equilibrium, we assume that: 

\begin{eqnarray}
\mu_{0}P_{0}'=p_{1}
\label{eqs:solov_assumpt_p}
\\
FF'=F_{0}
\label{eqs:solov_assumpt_f}
\end{eqnarray}
where $p_{1}$ and $F_{0}$ are constants\citep{solov_eq}. Furthermore, the plasma temperature and frequency of toroidal rotation are assumed to be constants $T_{0}$ and $\Omega_{0}$ respectively, i.e. $T=T_{0}$, $\Omega=\Omega_{0}$

The Grad-Shafranov equation Eq.(\ref{gs_modf_full}) is reduced to

\begin{eqnarray}
\Delta^{*}\psi=-p_{1}R^{2} \exp [M_{0}^{2}(\frac{R^{2}}{R_{0}^{2}}-1)]-F_{0}
\label{eqs:solov_gs1}
\end{eqnarray}
where $M_{0}= \frac{m_{i}R_{0}^{2}\Omega_{0}^{2}}{2T_{0}}$ denotes the Mach number at $R=R_{0}$.

The solution of Eq.(\ref{eqs:solov_gs1}) is of the form $\psi(R,Z)=\psi_{p}(R,Z) + \psi_{h}(R,Z)$, where $\psi_{p}$ is the particular solution and $\psi_{h}$ is the homogeneous solution\cite{solov_eq,solov_analy_solution1}.

\begin{eqnarray}
\psi_{h}=c_{1}+c_{2}R^{2}+c_{3}(R^{4}-4R^{2}Z^{2})+c_{4}[R^{2}ln(R)-Z^{2}]
\label{eqs:solov_solution_homog}
\end{eqnarray}
where these constants $c_{1},c_{2},c_{3},c_{4}$ are determined by boundary condition. Then, for a particular solution:

\begin{eqnarray}
\psi_{p}=-p_{1} \left( \frac{R_{0}^{2}}{2M_{0}^{2}} \right)^{2} \left\lbrace \exp \left[M_{0}^{2}\left(\frac{R^{2}}{R_{0}^{2}}-1\right)\right] - \frac{M_{0}^{2}}{R_{0}^{2}} \left( R^{2} - R_{0}^{2} \right) - 1 \right\rbrace-\frac{F_{0}}{2}Z^{2}
\label{eqs:solov_solution_part}
\end{eqnarray}

We obtain a new analytical solution of Grad-shafranov equation for the Solov'ev equilibrium with toroidal rotation:

\begin{eqnarray}
\psi=\psi_{p}+\psi_{h}=c_{1}+c_{2}R^{2}+c_{3}(R^{4}-4R^{2}Z^{2})+c_{4}[R^{2}ln(R)-Z^{2}] 
\label{eqs:solov_solution}
\\
\nonumber
-p_{1} \left( \frac{R_{0}^{2}}{2M_{0}^{2}} \right)^{2} \left\lbrace \exp \left[M_{0}^{2}\left(\frac{R^{2}}{R_{0}^{2}}-1\right)\right] - \frac{M_{0}^{2}}{R_{0}^{2}} \left( R^{2} - R_{0}^{2} \right) - 1 \right\rbrace-\frac{F_{0}}{2}Z^{2}
\end{eqnarray}
This solution reduces to the solution of static Solov'ev equilibrium when $\Omega_{0} \rightarrow 0$ or $M_{0} \rightarrow 0$.

\begin{equation}
\psi=\lim_{M_{0} \rightarrow 0} \psi_{h} + \psi_{p} = \psi_{h} + \lim_{M_{0} \rightarrow 0} \psi_{p}= \psi_{h} -p_{1} \frac{(R^{2} - R_{0}^{2})^{2}}{8} -\frac{F_{0}}{2}Z^{2}
\label{eqs:solov_solution_noflow}
\end{equation}
The above solution in Eq.(\ref{eqs:solov_solution_noflow}) was a specific case of the Grad-Shafranov equation solutions obtained before in Ref.\cite{solov_analy_solution1}. A similar solution of Solov'ev equilibrium with rigid toroidal rotation was recently obtained by Chu etal\cite{CHU_2018},

\begin{eqnarray}
\psi=\psi_{h}-p_{1} \left( \frac{R_{0}^{2}}{2M_{0}^{2}} \right)^{2} \left\lbrace \exp \left[M_{0}^{2}\left(\frac{R^{2}}{R_{0}^{2}}-1\right)\right] - \frac{M_{0}^{2}}{R_{0}^{2}} \left( R^{2} - R_{0}^{2} \right) - 1 \right\rbrace-\frac{F_{0}}{2}Z^{2}
\nonumber
\\
+\frac{1-2\beta_{pJ}}{16} \left[ (\frac{R^{2}}{R_{0}^{2}}-1)^{2} - \frac{4R^{2} Z^{2}}{R_{0}^{4}} \right]
\label{chu_analy}
\end{eqnarray}
where $\beta_{pJ}=-\frac{R_{0}^{2}p_{1}}{F_{0}+R_{0}^{2} p_{1}}$. The two solutions in Eqs.(\ref{eqs:solov_solution}) and (\ref{chu_analy}) differ only in the last term with the factor of $\frac{1-2\beta_{p,J}}{16}$.

\subsection{Maschke-Perrin Equilibrium}
\label{eqs_prev}

Another analytic solution of Eq.(\ref{intro:psi-direc}) was previously found based on the following assumptions\cite{M_P}:

\begin{eqnarray}
P=\frac{P_{0}}{R_{L}^{4}}(\psi-\psi_{1}) \exp \left(\gamma R^{2} \Omega^{2}/2R_{L}^{2} \right)
\label{eqs:mp_pres_asump}
\\
F^{2}=F_{0}^{2}+2\frac{M}{R_{L}^{2}}(\psi-\psi_{1})
\label{eqs:mp_f_asump}
\\
\frac{\omega^{2}}{\overline{R}T}={\rm constant}=\gamma \frac{\Omega^{2}}{R_{L}^{2}}
\end{eqnarray}
where $\gamma$ is the ratio of specific heats and $R_{L}$ $P_{0}$, $\psi_{1}$, $F_{0}$, $M$ are constants.

In case of $M=0$, the analytical solution takes the form

\begin{eqnarray}
\nonumber
\psi-\psi_{1}&=&C P_{0} \frac{R^{2}}{R_{L}^{2}}
\\
&+&P_{0} \left\{ \frac{(\epsilon_{a}-1)}{4} \left( \frac{Z^{2}}{R_{L}^{2}} - \frac{R^{2}}{4R_{L}^{2}} \right) \frac{R^{2}}{R_{L}^{2}} \right.
\label{eqs:mp_solution}
\\
\nonumber
&+& \left. \frac{1}{\gamma^{2}\Omega^{4}} \left[ 1+\frac{\gamma\Omega^{2}R^{2}}{2R_{L}^{2}} - \exp \left( \frac{\gamma \Omega^{2}R^{2}}{2R_{L}^{2}} \right) \right] \right\}
\end{eqnarray}
where $C=\frac{(\epsilon_{a}-1)}{8} r_{a}^{2} + \frac{1}{2\gamma \Omega^{2}} \left[ \exp \left( \frac{\gamma \Omega^{2}r_{a}^{2}}{2} - 1 \right) \right]$ is a constant, $\epsilon_{a}$ is a constant related to the ellipticity of the plasma cross-section, $r_{a}=R_{0}/R_{L}$ denotes the ratio between the position of magnetic axis $R_{0}$ and the chosen scale length $R_{L}$.

\section{Numerical algorithm}
\label{sec:numer}

NIMEQ solves the Grad-Shafranov equation in weak form using Galerkin formulation\cite{nimeq}. Defining one scalar field $\Lambda=\psi/R^{2}$, the Grad-Shafranov operator can be transformed into a divergence of a vector, $\Delta^{*} \psi = \nabla \cdot R^{2} \nabla \Lambda$. The scalar field $\Lambda$ can be spilt into two parts: $\Lambda_{0}$ and $\Lambda_{h}$ where $\Lambda_{0}$ satisfies the specified inhomogeneous boundary condition for $\Lambda$ and $\Lambda_{h}$ satisfies the boundary condition $\Lambda_{h}=0$. The $\Lambda_{h}$ is expended onto a series of $C^{0}$ spectral element basis functions $\Lambda_{h}=\sum_{i} \Lambda_{i} \alpha_{i}$.The weak form of Grad-Shafranov equation is obtained as:

\begin{eqnarray}
&&\sum_{i} \Lambda_{i} \int dVR^{2} \nabla \alpha_{i} \cdot \nabla \alpha_{j}=
\nonumber
\\
&&\int dV \left\{ FF^{'} + \mu_{0} R^{2} \left[ \frac{dP_{0}}{d\psi} + P_{0} \frac{m_{i} R_{0}^2 \Omega}{T} \left( \frac{R^{2}}{R_{0}^{2}}-1 \right) \frac{d\Omega}{d\psi} - P_{0} \frac{m_{i} R_{0}^{2} \Omega^{2}}{2T^{2}} \left( \frac{R^{2}}{R_{0}^{2}}-1 \right) \frac{dT}{d\psi} \right] \right.
\nonumber
\\
 && \left. \exp \left[ \frac{m_{i} \Omega^{2} R_{0}^{2}}{2T} \left( \frac{R^{2}}{R_{0}^{2}} -1 \right) \right]  \right\} \alpha_{j} - \int dV R^{2} \nabla \Lambda_{0} \cdot \nabla \alpha_{j}
\label{numer:weak_gs}
\end{eqnarray}

For compactness, Eq.(\ref{numer:weak_gs}) is written as $M\Lambda=Q$. The modified Picard iterations in Eq.(\ref{numer:iter_mat}), has been applied to solve the Grad-Shafranov equation in NIMEQ, where $\theta \in (0,1] $ denotes the relaxation parameter to achieve convergence. 

\begin{eqnarray}
M \Lambda^{n} = (1-\theta)M \Lambda^{n-1} + \theta Q^{n-1}
\label{numer:iter_mat}
\end{eqnarray}

After iteration, these equilibrium fields are calculated from the converged solution for $\Lambda$. The pressure, temperature, toroidal flow velocity $\vec{u}_{\phi}$ and $RB_{\phi}$ values are calculated from the prescribed $P_{0}(\psi)$, $F(\psi)$, $T(\psi)$, $\Omega(\psi)$ using the converged solution $\Lambda(R,Z)$ through Eq.(\ref{intro:pres_flow}) and $\vec{u}_{\phi}=R^{2} \Omega(\psi) \nabla \phi$. The poloidal magnetic field is expressed as Eq.(\ref{numer:bp}) in terms of $\Lambda$.

\begin{eqnarray}
\vec{B}_{p}=\frac{1}{R} \hat{e}_{\phi} \times (2R \hat{e}_{R} \Lambda + R^{2} \nabla \Lambda)
\label{numer:bp}
\end{eqnarray}
where $\hat{e}_{R}$ and $\hat{e}_{\phi}$ represent the unit vectors in the $R$ and $\phi$ directions respectively.

The poloidal current is calculated directly from the magnetic field through the relation $\vec{J}_{p}=-F^{'}\vec{B}_{p} / \mu_{0}$. And the toroidal current density is calculated using Eq.(\ref{numer:jphi})

\begin{eqnarray}
J_{\phi} =  \frac{1}{\mu_{0}R} \Delta^{*} \psi =R \frac{\partial P}{\partial \psi} + \frac{1}{\mu_{0} R} F \frac{\mathrm{d} F}{\mathrm{d} \psi}
\label{numer:jphi}
\end{eqnarray}

\section{Benchmark and Convergence}
\label{sec:bench_converg}

\begin{figure}[ht]
  \begin{center}
  \includegraphics[width=1\textwidth, height=0.5\textheight]{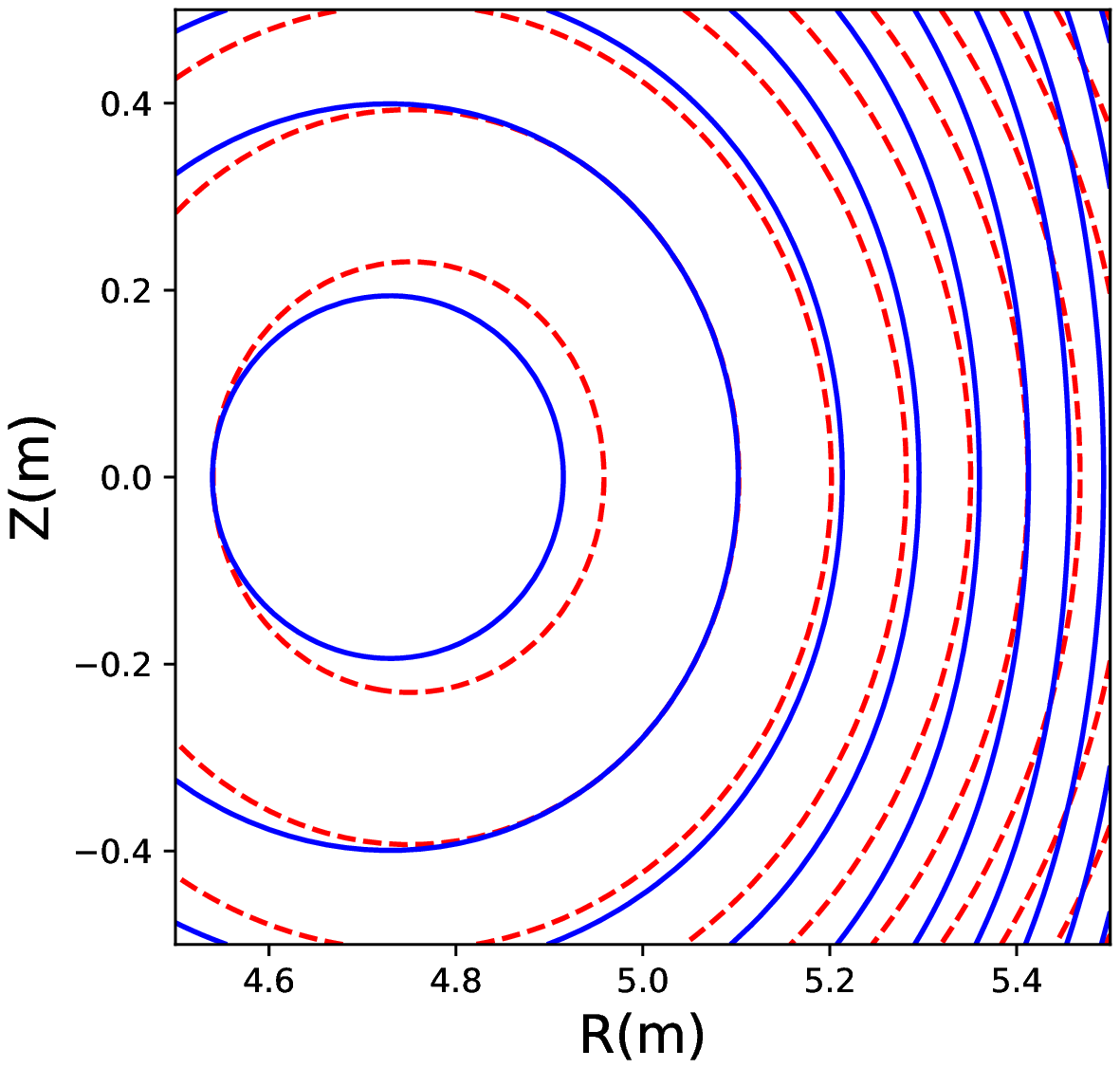}
  \end{center}
  \caption{The poloidal flux contour of the Solov'ev equilibrium with $p_{1}=-8.0 \times 10^{-2}$, $F_{0}=-20$ and $M_{0}=4$. The red dashed lines stand for the equilibrium without toroidal rotation. The blue solid lines stand for the equilibrium with toroidal rotation.}
\label{fig:con_bchmk}
\end{figure}

The analytic solutions in section \ref{eqs_new} are plotted in a domain of rectangular poloidal cross section with $4.5<R<5.5$ and $-0.5<Z<0.5$ (Fig.\ref{fig:con_bchmk}). Parameters are set as $p_{1} = -8.0 \times 10^{-2}$, $F_{0} = 20$, $M_{0} = 6.1$ and the poloidal flux along the boundary is prescribed using Eq.(\ref{eqs:solov_solution}), with $c_{1} = -104.1301$, $c_{2} = 10.6087$, $c_{3} = 0.0015$ and $c_{4} = -5.2103$. The equilibrium poloidal flux contours for Solov'ev equilibrium with toroidal rotation and without toroidal rotation presented in Fig.\ref{fig:con_bchmk} show modification induced by toroidal rotation.

\begin{figure}[ht]
  \begin{center}
  \includegraphics[width=1\textwidth, height=0.5\textheight]{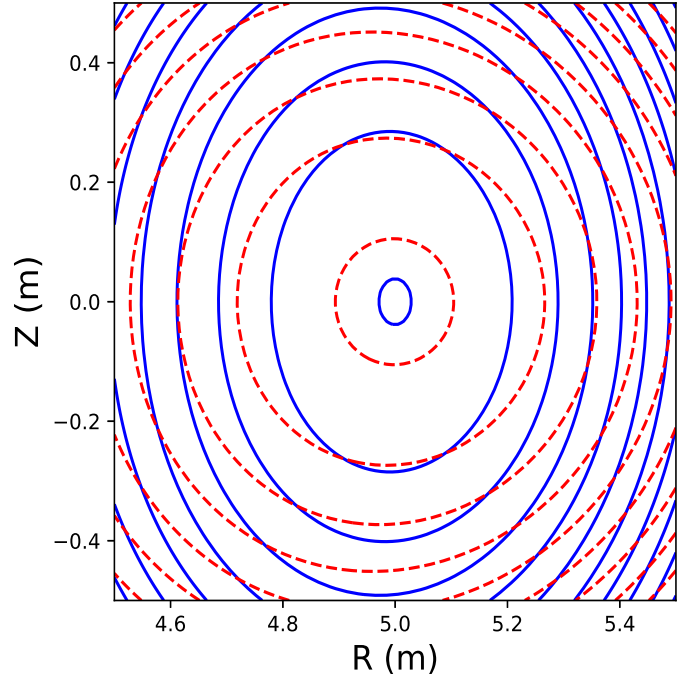}
  \end{center}
  \caption{The poloidal flux contour of the Maschke and Perrin's equilibrium with $P_{0}=-8.0 \times 10^{-3} \times R_{L}^{4}$ and $\Omega=5.0 \times 10^{4}$. The red dashed lines stand for the equilibrium without toroidal rotation. The blue solid lines stand for the equilibrium with toroidal rotation.}
\label{fig:con_mp}
\end{figure}

Similarly, the equilibrium poloidal flux contours for Maschke and Perrin's equilibrium in section \ref{eqs_prev} are plotted in a domain of rectangular poloidal cross section with $4.5 \leq R \leq 5.5$ and $-0.5 \leq Z \leq 0.5$ (Fig.\ref{fig:con_mp}). In this case, we choose $\epsilon_{a}=0$, $\gamma=5/3$, $R_{0}=R_{L}=5.0$, $P_{0}=-0.1 \times R_{L}^{4}$, $\Omega=3.0 \times 10^{5}$, $\psi_{0}=0$ and $\varepsilon_{a}=0$. Distortion of flux surfaces due to toroidal rotation is also apparent.

Both benchmark and convergence studies are performed for Solov'ev equilibrium and Maschke and Perrin's equilibrium by comparing the numerical and analytical solutions. The numerical error of equilibrium poloidal flux is defined as $E_{n}=\sqrt{\sum(\psi_{n}-\psi_{a})^{2}/ \sum \psi_{a}^{2}}$, where $\psi_{n}$ is the numerical solution from the extended NIMEQ and $\psi_{a}$ is the analytic solution from Eq.(\ref{eqs:solov_solution}) and Eq.(\ref{eqs:mp_solution}). And the summation is performed over all of the finite-element nodes. 

Two methods, i.e. h-refinement and p-refinement, are applied to checking the convergence of the extended NIMEQ in both equilibria. In the p-refinement method, the polynomial degree of each element is increased whereas the number of elements is kept constant. H-refinement maintains the polynomial degree of the elements while increasing the number of elements. The decaying rate of the error for a smooth solution of a second order differential equation is bounded by the asymptotic rate of convergence $h^{(p+1)}$ for sufficiently smooth solutions, where $h$ is a characteristic element length of calculation region and p is the polynomial degree\citep{convg_geo}.

\begin{figure}[ht]
  \begin{center}
  \includegraphics[width=1\textwidth, height=0.5\textheight]{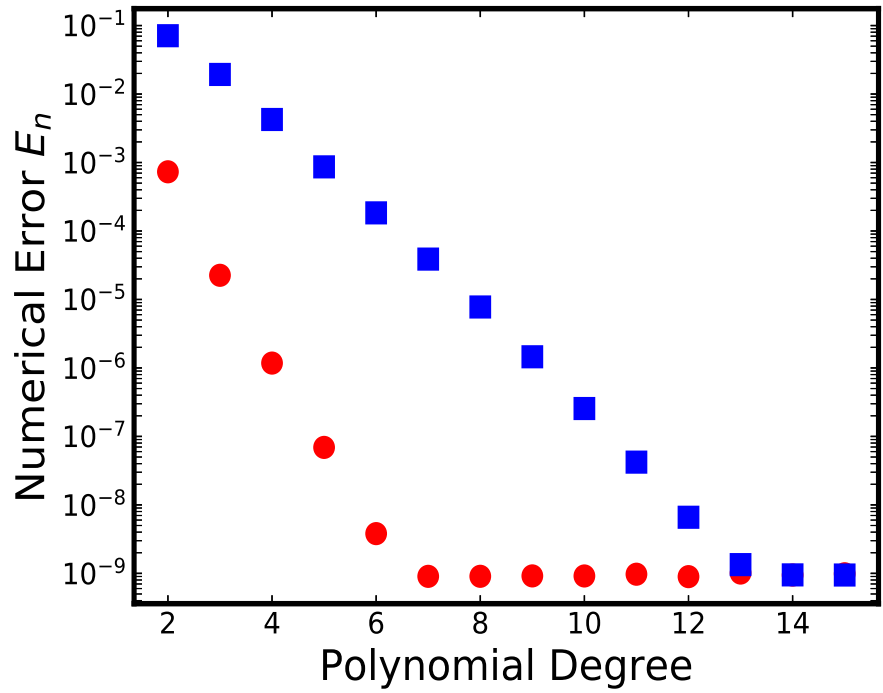}
  \end{center}
  \caption{The numerical error $E_{n}$ of $\psi$ as a function of the element polynomial degree for $2 \times 2$ element mesh ({\color{blue}{$\blacksquare$}}) and $10 \times 10$ element mesh ({\large{\color{red}{$\bullet$}}}) in the case of Solov'ev equilibrium with toroidal rotation.}
\label{fig:p_convergence}
\end{figure}

\begin{figure}[ht]
  \begin{center}
  \includegraphics[width=1\textwidth, height=0.5\textheight]{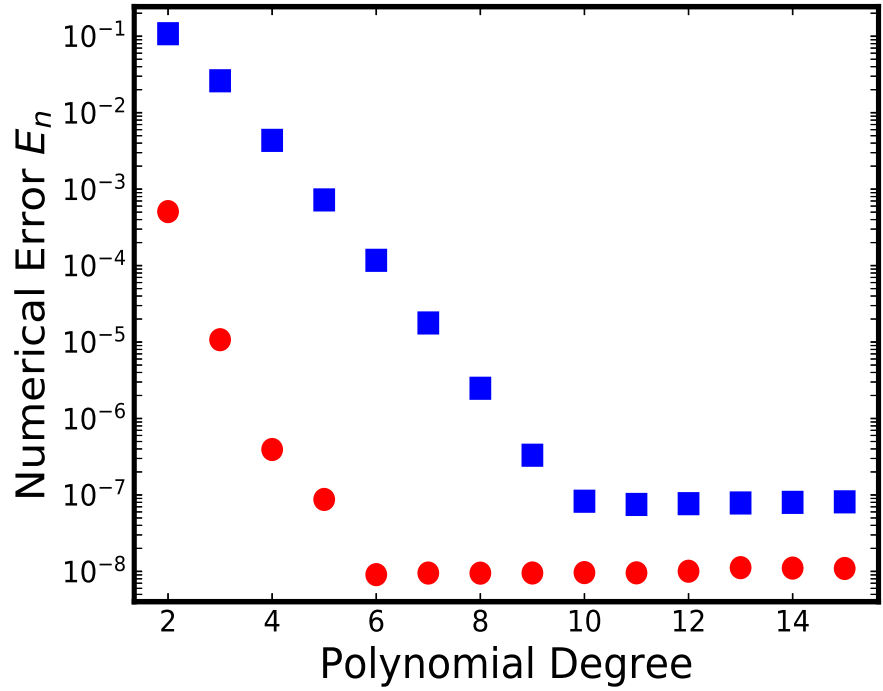}
  \end{center}
  \caption{The numerical error $E_{n}$ of $\psi$ as a function of the element polynomial degree for $2 \times 2$ element mesh ({\color{blue}{$\blacksquare$}}) and $10 \times 10$ element mesh ({\large{\color{red}{$\bullet$}}}) in the case of Maschke and Perrin's equilibrium.}
\label{fig:p_convergence_mp}
\end{figure}

We use meshes with equal numbers of elements in the radial and vertical directions. In the p-refinement study, the polynomial degree of elements is scanned from 2 to 15 when keeping the $2 \times 2$ and $10 \times 10$ element meshes fixed for both equilibria. In both equilibrium cases, the numerical errors decay linearly to a minimum value, which indicates geometric convergence in Fig.\ref{fig:p_convergence} and Fig.\ref{fig:p_convergence_mp} \citep{convg_geo}. The numerical error in $10 \times 10$ element meshes decays faster than that in $2 \times 2$ element meshes in both equilibria.

\begin{figure}[ht]
  \begin{center}
  \includegraphics[width=1\textwidth, height=0.5\textheight]{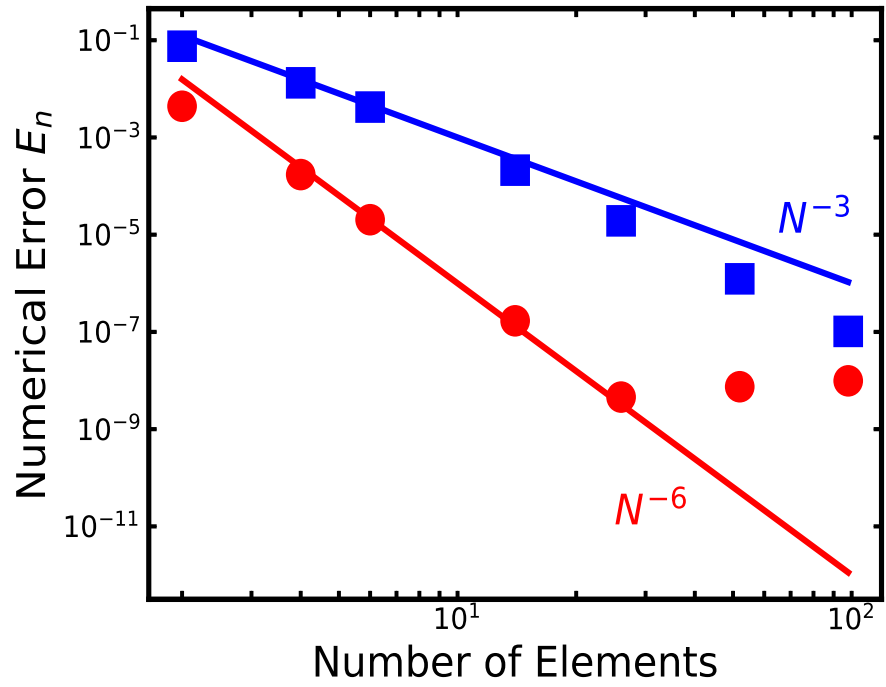}
  \end{center}
  \caption{The numerical error $E_{n}$ of $\psi$ as a function of the element numbers for 2nd order elements ({\color{blue}{$\blacksquare$}}) and 4th order elements ({\large{\color{red}{$\bullet$}}}) in the case of the Solov'ev equilibrium with toroidal rotation.}
\label{fig:h_convergence}
\end{figure}

\begin{figure}[ht]
  \begin{center}
  \includegraphics[width=1\textwidth, height=0.5\textheight]{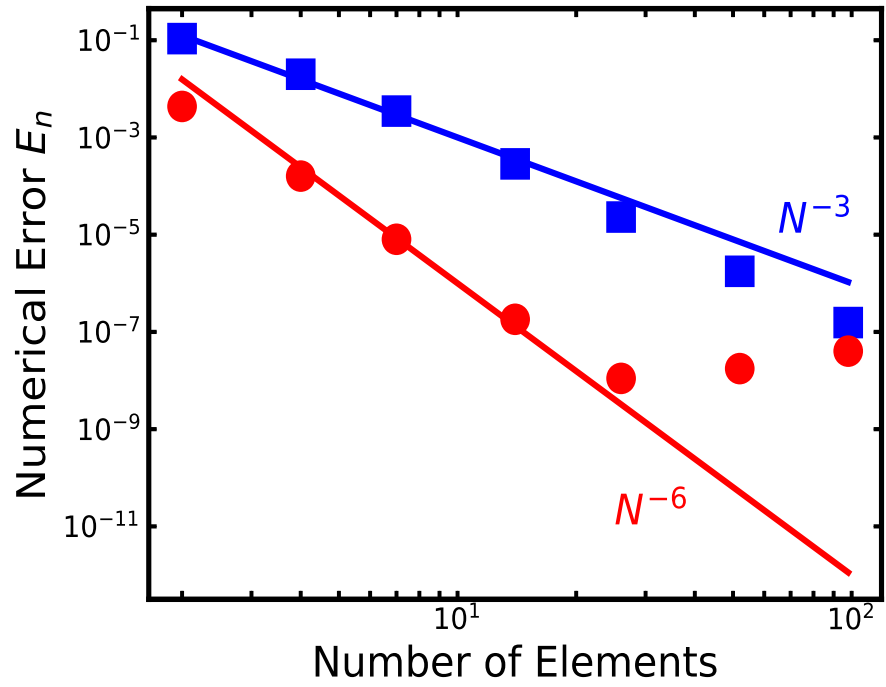}
  \end{center}
  \caption{The numerical error $E_{n}$ of $\psi$ as a function of the element numbers for 2nd order elements ({\color{blue}{$\blacksquare$}}) and 4th order elements ({\large{\color{red}{$\bullet$}}}) in the case of Maschke and Perrin's equilibrium.}
\label{fig:h_convergence_mp}
\end{figure}

In h-refinement study, the number of elements are scanned from 4 to 94 when polynomial degree of elements keeps 2 and 4. In h-refinement studies of both equilibrium cases, the numerical errors decay linearly to a minimum value, indicating algebraic convergence in Fig.\ref{fig:h_convergence} and Fig.\ref{fig:h_convergence_mp}\citep{convg_geo}. The decay rate of numerical erros with polynomial degree fixed 4 is larger than that with polynomial degree fixed 2. The blue lines in both figures stand for the scaling $N^{-3}$ and $N^{-6}$  fitted from the decaying numerical errors, where $N$ denotes the number of elements. Both figures show that the decay rates of numerical error in $\psi$ are between $p+1$ and $p+2$. 

\begin{figure}[ht]
  \begin{center}
  \includegraphics[width=1\textwidth, height=0.5\textheight]{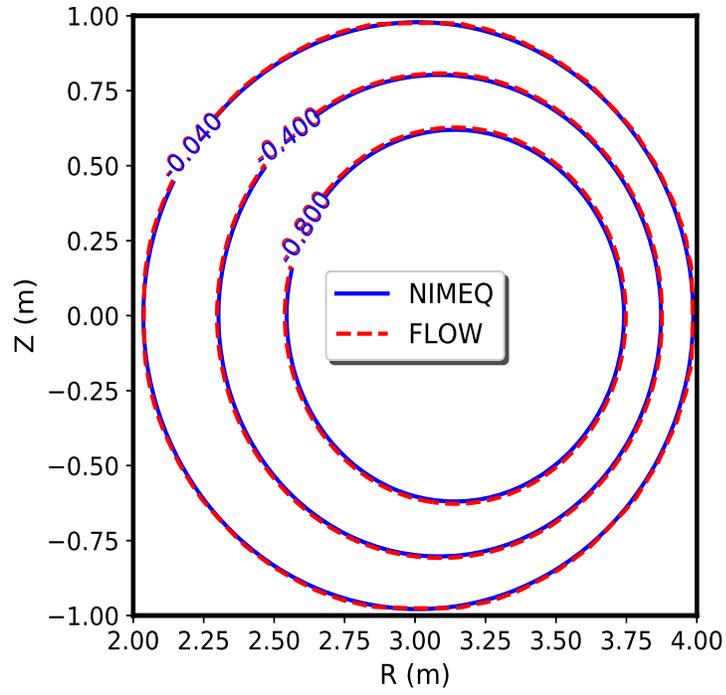}
  \end{center}
  \caption{The comparison between $\psi$ from the extended NIMEQ and $\psi$ from  FLOW code, with $P_{\rm{open}}=1.0 \times 10^{-3}$, $P_{1}=0.8$, $P_{2}=0.2$, $n_{\rm{axis}}=8.0 \times 10^{19}$ and $F=4.0$. The red dashed lines stand for the $\psi$ from FLOW code and blue solid lines denote the $\psi$ from the extended NIMEQ.}
\label{fig:compare_nimeq_FLOW}
\end{figure}
\begin{figure}[ht]
  \begin{center}
  \includegraphics[width=1\textwidth, height=0.5\textheight]{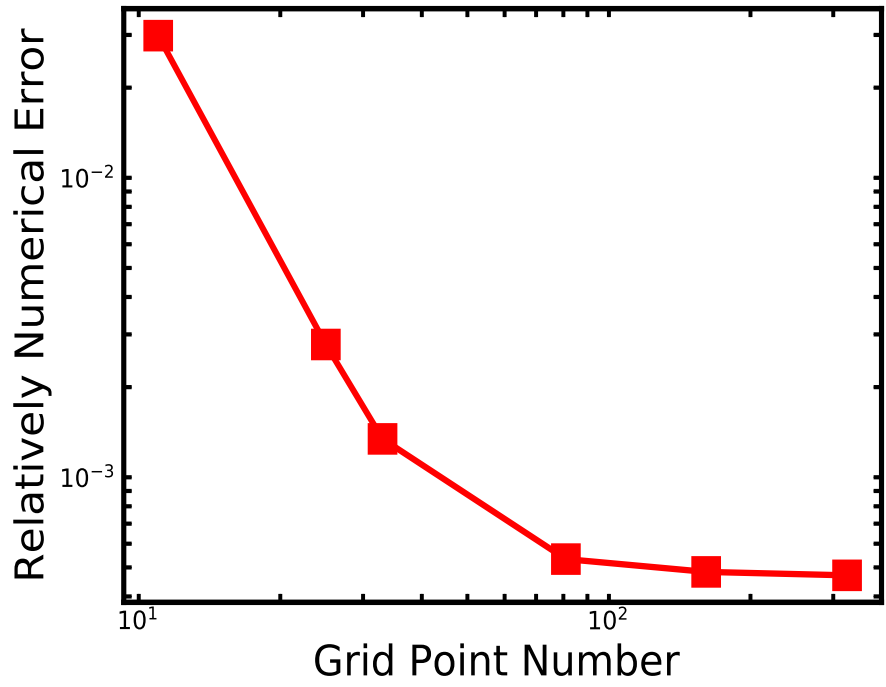}
  \end{center}
  \caption{The relatively error as a function of grid point number.}
\label{fig:convergence_nimeq_FLOW}
\end{figure}

FLOW is a finite difference code, which solves the Bernoulli-Grad-Shafranov equations for tokamak the equilibriums with flow\cite{FLOW_code}. The extended NIMEQ is benchmarked with the FLOW code here. The comparison is performed in a poloidal domain of $2.0 \leq R \leq 4$ and $-1.0 \leq Z \leq 1.0$. The $F$ and $\Omega$ are chosen as constants. And the pressure profile is specified as one quadratic function of the normalized $\psi$, $P_{0}(\psi) = P_{\rm{open}} +P_{1} (1-\psi) + 4 P_{2} \psi (\psi-1)$. The number density profile is similar to the pressure profile, since $n(\psi)=\frac{P_{0}(\psi)}{P_{0}(0)} n_{\rm{axis}}$, where $P_{0}(0)$ and $n_{\rm{axis}}$ denote the pressure and number density on magnetic axis. This number density profile is thus chosen so as to obatain a consatnt temperature. The Mach number is constant and equals 0.3. The overlay of $\psi$ form the extended NIMEQ and FLOW is shown in Fig.\ref{fig:compare_nimeq_FLOW}. For comparison, the relatively numerical error is defined as $\sqrt{ \sum (\psi_{\rm{FLOW}}-\psi_{\rm{NIMEQ}})^{2} / \sum \psi_{\rm{NIMEQ}}^{2}}$ where $\psi_{\rm{NIMEQ}}$ denotes the numerical solution from NIMEQ and $\psi_{\rm{FLOW}}$ denotes the numerical solution from FLOW. Because the computation grids are different in NIMEQ and FLOW, the bi-cubic spline interpolation is applied to calculation of relatively numerical error. The relative numerical error decreases with the computation grid point number (Fig.\ref{fig:convergence_nimeq_FLOW}).

\section{Conclusion and discussion}
\label{sec:conclusion}

We have extended NIMEQ by solving the modified Grad-Shafranov equation that self-consistently  takes into account of the effects of toroidal rotation. A new analytic solution to the modified Grad-Shafranov equation is obtained for the Solov'ev equilibrium in presence of a rigid toroidal rotation. Both the new analytical solution and the Maschke-Perrin equilibrium are used in benchmark and convergence studies. High accuracy solution with numerical error to the order of $10^{-10}$ or smaller is achieved. The extended NIMEQ is also successfully benchmarked with the FLOW code.

Next we plan to extend the modified Grad-Shafranov equation to include free boundary condition, and to study the effects on equilibrium profiles due to toroidal rotation. Meanwhile, the poloidal flow can be also included in the extended NIMEQ to calculate equilibrium in presence of arbitrary flows.

\section*{Acknowledgments}
This work was supported by the Fundamental Research Funds for the Central Universities at Huazhong University of Science and Technology Grant No. 2019kfyXJJS193, the National Natural Science Foundation of China Grant No. 11775221, the National Magnetic Confinement Fusion Science Program of China Grant No. 2015GB101004, and U.S. Department of Energy Grant Nos. DE-FG02-86ER53218 and DE-SC0018001. This research used the computing resources from the Supercomputing Center of University of Science and Technology of China.


\newpage

\newpage

\newpage

\newpage

\newpage

\newpage

\newpage

\newpage

\newpage

\end{document}